# Integrating Endothelial-Derived Hyperpolarizing Signaling into a Multitarget Therapeutic Strategy for Microvascular Disease

Paolo Madeddu, MD;[1] Styliani Goulopoulou, PhD;[2] Dave Wambeke, BS.[3]

1. Experimental Cardiovascular Medicine, Bristol Medical School, Translational Health Sciences, University of Bristol, Bristol, United Kingdom.
2. Lawrence D. Longo, MD Center for Perinatal Biology, Dept. Basic Sciences, Dept. Gynecology & Obstetrics, Loma Linda University, Loma Linda, California, USA.
3. DiaMedica Therapeutics, Minnetonka, Minnesota, USA.

Corresponding author:
Prof. Paolo Madeddu.
University of Bristol,
Upper Maudlin Street, Bristol, BS2 8HW, UK.
Email: mdprm@bristol.ac.uk.

**Abstract**

Endothelial cells release various vasorelaxing molecules, such as nitric oxide (NO) and prostacyclin (PGI$_2$), along with less clearly defined factor(s) that induce hyperpolarization of vascular smooth muscle cells through the opening of calcium-sensitive potassium channels. Potassium channel-dependent vasorelaxation is prevalent in microvessels and can partially compensate for deficiencies in other vasodilatory mechanisms. Enhancing this backup vasorelaxant mechanism may aid the treatment of microvascular disorders, such as cerebral small vessel disease and preeclampsia, a pregnancy-specific hypertensive syndrome, which is characterized by systemic endothelial dysfunction. The development of pharmacological potassium channel openers has encountered significant challenges, including issues of specificity, safety concerns, and off-target effects. This study critically evaluates the advantages and drawbacks of integrating hyperpolarization into a holistic vasorelaxant strategy for managing of ischemic disease through single or combination drug therapies.

**Introduction**

The endothelium has historically been regarded as an inert layer encasing the lumen of blood arteries, primarily functioning as a barrier between circulating blood and the outer layers of the vascular wall. The identification of endothelium-derived relaxing factors, nitric oxide (NO) and prostacyclin ($PGI_2$), marked a significant shift in perspective.[1-4] Subsequent observations indicated that agonist-induced relaxation persists in part after the inhibition of NO and $PGI_2$ signaling, implying the presence of an alternative vasorelaxant mechanism.[5,6] Endothelium-dependent relaxation, independent of NO and $PGI_2$ production and occurring without any significant increase in intracellular levels of cyclic nucleotides (cGMP and cAMP), can be induced by diffusible factors of diverse origins, including arachidonic acid metabolites, gaseous mediators, and reactive oxygen species. The hyperpolarization of vascular smooth muscle cells is a common mechanism among these processes, leading to the designation of each factor as "endothelium-derived hyperpolarizing factor" or "EDHF" at different points.

Another pathway, described as "endothelium-dependent hyperpolarization" (EDH), which does not inherently involve a diffusible factor, also significantly contributes to the endothelium-dependent regulation of vascular tone. The pathway necessitates the activation of endothelial calcium ($Ca^{2+}$)-activated potassium channels (KCa channels) to generate hyperpolarization, which then spreads between endothelial cells, passing into the muscle layers to initiate vasodilation. EDH physiologically promotes conducted (or ascending) vasodilation, essential for ensuring adequate perfusion to meet the metabolic demands of peripheral tissues. Significantly, vasodilatation also regulates blood pressure by affecting peripheral vascular resistance.

The evolution of complex organisms involved the accrual of redundant vasorelaxant pathways, which ensured superior functional resilience against environmental disturbances. Additionally, the expression of various relaxant pathways evolved according to finely tuned organ- (organotypic) and vessel-specific (angiotypic) programs. Since the inception of modern vascular biology, EDH has been recognized as a mechanism that explicitly regulates the microcirculation.[7,8] In a key 'Current Awareness' paper published in Trends in Pharmacological Sciences, Taylor and Weston posited that hyperpolarization induces vasodilation by shutting voltage-dependent Ca2+ channels. They asserted, "This process is likely crucial in small arterioles that rely on Ca2+ influx during contraction."

This concise review article highlights current evidence for hyperpolarization as a vital compensatory mechanism that safeguards against microvascular hypoperfusion in conditions of diminished or uncoupled NO production. The second section underscores the potential for targeting hyperpolarization-dependent vasorelaxation to address the unmet therapeutic needs of microvascular pathologies, such as cerebral small vessel disease, a common cause of stroke and dementia, and preeclampsia, a complication of pregnancy associated with serious risks to both mother and offspring. . A proposed solution involves a multi-target product that serves as a master key to address the mechanisms underlying unbalanced microvascular constriction and hypoperfusion.

**Elements of hyperpolarization-mediated vascular relaxation**

Potassium ($K^+$) channels comprise families that are constitutively active or activated by an increase in intracellular $Ca^{2+}$ ions or by membrane depolarization. The nomenclature and functional description of $K^+$ channels have been continuously updated in recent years due to the discovery of several new families.[9-12] The Nomenclature Committee of the International Union of Basic and Clinical Pharmacology (NC-IUPHAR) Subcommittees on K+ channels have proposed a standardized nomenclature, categorizing groups according to gene family and the structural composition of 6, 4, or 2 transmembrane domains.[12]

*$Ca^{2+}$ activated $K^+$ channels (KCa):* This family comprises three subtypes, each exhibiting distinct physiological or pathological roles in various cells and tissues: big conductance (BKCa), intermediate conductance (IKCa), and small conductance (SKCa) channels.[11] Both BKCa and IKCa channel families consist of a single member (KCa1.1 and KCa3.1, respectively), while the SKCa family includes three subtypes (KCa2.1, 2.2, and 2.3). The KCa2.X and KCa3.1 channels share analogous sequences but demonstrate structural differences from the KCa1.1 channel.[10-13] In the majority of arterial beds, endothelium-derived hyperpolarization entails the activation of both small KCa2.3 and intermediate KCa3.1 channels that are typically confined to the endothelium. BKCa channels are mainly expressed in vascular smooth muscle cells and the inner mitochondrial membrane of cardiomyocytes. These channels are distinguished by their pronounced $K^+$ selectivity, substantial single-channel conductance (approximately 10–20 times greater than other $K^+$ channels), and a remarkable capacity for dual activation by two separate physiological stimuli: membrane depolarization and localized elevations in intracellular $Ca^{2+}$.[13] The activation of BKCa channels, characterized by their very large unitary conductance, induces a rapid efflux of $K^+$, culminating in membrane hyperpolarization. In the vasculature, BKCa function as a negative feedback mechanism for vasocontraction, contributing to physiologic vasodilation.[14] In line with these essential physiological roles, both dysfunction and aberrant expression (loss or gain of function) of BKCa channels can adversely affect vascular function, contributing to the pathogenesis of various diseases, including hypertension and diabetic complications.[15]

*The inward-rectifier $K^+$ channel family:* This family comprises the constitutively active inward-rectifier $K^+$ channels (Kir2.x), the G-protein-activated inward-rectifier $K^+$ channels (Kir3.x), and the ATP-sensitive $K^+$ channels (Kir6.x), which associate with sulphonylurea receptors. The pore-forming subunits form tetramers, and within subfamilies, heteromeric channels may occur (e.g., Kir3.2 with Kir3.3). Kir channels are considered the most prominent $K^+$ channels in endothelial cells, but are also expressed by capillary pericytes. They are activated by laminar flow, G protein-coupled receptor (GPCR) agonists, and $K^+$ efflux from neighbouring cells.[16] Kir channels enable a large $K^+$ influx but little outward current outflow under physiological conditions.[16] This causes additional $K^+$ efflux and an enhancement of hyperpolarization, which can be eventually transmitted through the endothelial layer to dilate upstream feeding arteries.[17]

*Voltage-gated $K^+$ channels*: This evolutionarily conserved family comprises genes and is classified into twelve subfamilies. KV channels exhibit extensive distributions within the nervous system and various other tissues, including the vasculature. These channels are activated at more negative membrane voltages than KCa in vascular smooth muscle cells of small arterioles, thus conspicuously contributing to vascular tone control.[18] The subfamily of Kv7 channels includes

Kv7.1 through Kv7.5 (KCNQ1-5). In the brain vasculature, Kv7 channels play a crucial role in regulating vascular tone and maintaining the integrity of the blood-brain barrier.[19,20] They can also contribute to vasorelaxation of the coronary arteries and human chorionic plate arteries.[21,22]

**Hyperpolarization crosstalk between endothelial cells and vascular smooth muscle cells**

The activation of SKCa and IKCa channels results in the hyperpolarization of endothelial cells, where $K^+$ efflux through these channels functions as a diffusible EDH factor. This current transmits from the endothelium to smooth muscle cells through myoendothelial gap junctions (MEGJs) situated within a myoendothelial projection (MEP), thereby inhibiting $Ca^{2+}$ influx via voltage-gated $Ca^{2+}$ channels (VGCCs).[23] The efflux of $K^+$ and/or hyperpolarization activates sodium ($Na^{+)}$//$K^+$-ATPase and Kir channels, leading to additional $K^+$ efflux and an enhancement of hyperpolarization. Decreased $Ca^{2+}$ influx into smooth muscle cells impairs myosin phosphorylation by $Ca^{2+}$-dependent myosin light chain kinase (MLCK), reducing vascular resistance.[24] Hyperpolarizing currents induced by agonists such as acetylcholine and bradykinin can swiftly propagate bidirectionally along the endothelium, resulting in anterograde and retrograde vasorelaxation. The axial orientation of endothelial cells facilitates these phenomena, as they typically extend over 10 to 20 smooth muscle cells.

**The relationship between hyperpolarization and other vasodilatory mechanisms**

Conduit elastic arteries, such as the aorta, carotid arteries, and major cerebral arteries, possess a large diameter that facilitates the transport of significant blood volumes with minimal resistance. Small diameter arteries, including penetrating arterioles in the brain and decidual spiral arteries in the endometrium in early pregnancy, play a critical role in regulating local tissue perfusion. NO signalling and EDH induced through KCa channel opening demonstrate angiotypic variations, with the former signaling mechanism predominating in conductance arteries and the latter in arterioles.

Importantly, NO and EDH exhibit a reciprocal counter-regulatory relationship. Exogenous NO administration at concentrations similar to those seen after stimulation with endothelium-dependent agonists reduces EDH-mediated vasorelaxation in rabbit carotid and pig coronary arteries.[25] The inhibitory effect is likely due to disrupting the synthesis and/or release of EDH-like factors. EDH-dependent vasorelaxation is notably enhanced when eNOS is inhibited or when NO production is impaired. Small arteries display eNOS within caveolae, which are specialized invaginations of the lipid plasma membrane, creating a docking site for structural complexation with the hemoglobin α-chain and cytochrome B5 reductase 3. The reduction of hemoglobin α-chain increases NO scavenging, thereby limiting its local bioavailability. This NO microdomain is typically deficient in conduit arteries, thus accounting for the increased NO bioavailability in this region.[26,27]

Risk factors and diseases affect NO generation through multiple pathways. The intracellular concentration of L-arginine serves as the rate-limiting factor in eNOS-mediated NO release. The enzyme arginosuccinate lyase facilitates the conversion of citrulline to arginine, thereby maintaining stable cellular concentrations of L-arginine. In pathological conditions such as hypertension, diabetes, and hypercholesterolemia, eNOS may undergo uncoupling. Uncoupled eNOS generates superoxide radicals at the expense of NO, thereby worsening oxidative stress and endothelial dysfunction. Uncoupling may result from L-arginine deficiency or its cofactor tetrahydrobiopterin (BH4).[28] Despite the high cellular concentrations of L-arginine, its excessive catabolism by degrading enzymes such as arginase, or the accumulation of asymmetric dimethylarginine (ADMA), can lead to deficiency or functional impairment. Excessive oxidation induced by radicals may result in a deficiency of BH4. Other uncoupling scenarios involve the failure of NOS dimerization and the glutathionylation of eNOS, which impairs endothelium-dependent vasodilation.[29,30] Uncoupling can be experimentally replicated using competitive NOS activity inhibitors, such as methylated and nitrated derivatives of L-arginine (e.g., L-NAME and L-NMMA).

The EDH signalling pathway may contribute to or compensate for endothelial dysfunction, depending on the specific cardiovascular condition, suggesting that this mechanism functions as a supplementary vasodilatory mechanism.[31] The following sections examine the critical function of EDH in regulating two microvascular compartments: the smaller cerebral vessels and the uteroplacental arteries. The brain and uteroplacental vascular unit exhibit significant similarities, particularly in their potent vasorelaxant and remodelling properties. Microvascular endothelial cells are involved in the structure of a barrier in both organs that safeguards against potential harm from circulating substances.

**The role of EDH signalling in the regulation of small cerebral vessel relaxation**

Cerebral arteries and arterioles modulate their diameter in response to pressure fluctuations, thereby autoregulating cerebral blood flow and protecting microcirculation from harmful hydrostatic pressure effects. The smaller vessels of the brain include cortical and deep penetrating arteries. Cortical arteries supply blood to the cerebral cortex. In contrast, deep penetrating arteries nourish interior regions such as the basal ganglia and thalamus. Smaller vessels demonstrate a greater basal tone compared to larger cerebral vessels. Cerebral blood vessels can dilate significantly, adjusting blood flow in response to brain activity, a neurovascular coupling phenomenon, to meet increased demands for oxygen and nutrients.

Consensus exists that NO-induced vasorelaxation diminishes throughout the cerebrovascular system, in contrast to EDH-dependent vasorelaxation, which is heightened in the brain microvasculature. Here, neural-endothelial coupling functions through hyperpolarization to maintain perfusion in physiological and pathological contexts. Capillary endothelial cells detect brain activity and transmit it to upstream arterioles as an electrical signal for vasodilation. Extracellular $K^+$, a consequence of neuronal activity, prompts the initiation of this signal. This activates inward-rectifier $K^+$ (KIR2.1) channels in capillary endothelial cells, leading to a rapid hyperpolarization that causes retrograde vasodilation, ultimately influencing upstream arterioles and pial arteries and increasing blood flow to the capillary bed. [32,33]

**Alterations of $K^+$ channels in microvascular cerebral disease**

Small vessel disease constitutes 25% of stroke occurrences and is associated with 45% of dementia cases. The prevalence increases with age, affecting about 5% of individuals aged 50 and nearly 100% of those over 90.[34] Anatomical features encompass wall thickening, disorganization, breakdown, and perivascular edema, which disrupt this connection and may result in microbleeds and lacunar infarcts. This remodelling is associated with and may precede the disruption of the blood-brain barrier (BBB). The semi-permeable barrier encasing the brain's microvasculature, distinguished by numerous tight junctions, governs the influx and efflux of circulating substances. It also protects the central nervous system from toxins and infections.[35]

A study investigating the role of SKCa and IKCa channels in myogenic tone under baseline conditions and following ischemia/reperfusion indicated that channel activity diminishes the basal tone of penetrating arteries without affecting large cerebral arteries. In cerebral ischemia/reperfusion, there is an increase in thromboxane $A_2$, a prothrombotic agent generated by activated platelets, and peroxynitrite, a byproduct of NO and superoxide. They generate opposing effects on endothelium-derived hyperpolarization. The activation of thromboxane and prostaglandin receptors reduces hyperpolarization, while peroxynitrite increases it. The observation that small brain arteries maintain or possibly enhance their reactivity to hyperpolarization factors following middle cerebral artery closure and reperfusion suggests a potential role for this vasodilator mechanism in eNOS inhibition after brain ischemia/reperfusion.[36]

KCa channel dysfunction may contribute to microvascular dementia. APP23 mice reportedly exhibit an approximately sevenfold elevation in amyloid precursor protein, comparable to levels found in Alzheimer's disease patients.[37] Pial arteries from APP23 mice exhibit heightened pressure-induced (myogenic) constriction, accompanied by a significant decrease in ryanodine receptor-mediated local $Ca^{2+}$ release in arterial smooth muscle cells. This resulted in a reduction of the activity of BKCa channels. The ability of the endothelial Kir2.1 channel to facilitate vasodilation was similarly compromised. The acute administration of amyloid-β 1-40 peptide to cerebral arteries from wild-type mice effectively replicated the BKCa channel dysfunction seen in APP23 mice, while not affecting Kir2.1 function.[37] Recent discoveries reveal that the voltage-gated Kv7 channels are crucial in modulating the activity of cardiovascular-related neurons, and their expression is diminished in hypertension. [19]

Conversely, the EDH signalling may negatively affect the permeability of the BBB endothelium. KCa3.1 current and channel protein expression have been observed in the endothelial cells of the BBB in both bovine and human subjects.[38] Cerebral edema occurs in the early phases of ischemic stroke due to the transendothelial release of $Na^+$, $Cl^-$, and water into the brain. Evidence suggests that the expression and activity of the KCa3.1 channel influence the behaviour of BBB endothelial cells. Thus, pharmacological inhibition of KCa3.1 channels may provide a practical therapeutic approach for reducing cerebral edema in the first three hours following ischemic stroke.[38] Another study demonstrated the expression and functional role of Kv7 channels in the BBB. Activation of Kv7 channels diminished endothelial cell permeability under both normal and pathological settings by inducing hyperpolarization of the cell membrane and reinforcing tight junctions. Consequently, the activation of endothelial Kv7 channels has been suggested as a practical approach for treating diseases associated with BBB impairment [20]

The data underscore the essential role of EDH as a primary mechanism of cerebral vasodilation. KCa and voltage-gated Kv7 channel activity protects against cerebral ischemia and dementia by compensating for the dysfunction of other vasorelaxant pathways, thereby partially preserving brain perfusion after ischemic injury and arterial hypertension. KCa channel-mediated hyperpolarization of the BBB endothelium may be associated with increased microvascular permeability, potentially contributing to cerebral edema after an acute ischemic stroke. The Kv7 channels may exert an opposite action.

**EDH signaling in the uteroplacental circulation during pregnancy**

The uterine spiral arteries, which are the terminal branches of the uterine artery, function as conduits from the myometrium to the decidua, delivering oxygen and nutrients to the developing placenta. During pregnancy, the uterine vasculature undergoes expansive hypertrophic remodelling and functional changes.[39,40] These adaptations support the metabolic needs of a growing uterus and the development of the fetoplacental unit. As a result, uterine blood flow increases 10-to 20-fold compared to the non-pregnant state, attaining 800-1000 mL/min by term, which is approximately 20% of cardiac output.[41]

An angiotypic program characterizes the anatomical and functional adaptation of the uterine circulation from large to small uterine vessels. Their decidual segments, consisting of specialized mucosal lining of the endometrium, undergo endovascular trophoblast migration, replacing the arterial wall's muscular and elastic layers with a dense, fibrinoid matrix.[42] This remodelling lowers downstream resistance and is associated with upstream vasorelaxation of the radial and arcuate uterine arteries.

Functionally, the uterine arteries show reduced sensitivity to vasoconstrictor adrenergic agonists and enhanced relaxation responses to dilatory agents. [43] These changes reflect increased local production of NO and $PG_{I2}$, along with enhanced EDH-mediated responses. EDH has a prevalent contribution to endothelium-dependent dilation in distal human myometrial arteries from non-pregnant and late pregnant women, as indicated by the marked relaxation in response to bradykinin or placental growth factor (PlGF) in the presence of NO and cyclooxygenase inhibition.[44-46] Similarly, EDH has a significant contribution to acetylcholine-induced relaxation in rat radial uterine arteries, confirming that the role of EDH in maternal uterine vascular tone increases in distal uterine vessels.[47] The predominant roles of NO, $PGI_2$, and EDH vary during gestation. In late pregnancy, EDH-mediated vasodilation and the associated reduction in smooth muscle intracellular $Ca^{2+}$ concentrations are reportedly enhanced in rat uterine resistance arteries, coinciding with higher endothelial $Ca^{2+}$ concentrations and activation of SKCa and IKCa channels.[48-50]

Various modulators regulate EDH signalling by changing SKCa channel expression. Estrogen is a significant activator of these channels in the uterine circulation throughout gestation. Expression of the SKCa2.3 gene (*KCNN3*) is modulated through interactions with estrogen receptor α, and the subsequent upregulation is associated with improved channel stoichiometry, thereby increasing $Ca^{2+}$ sensitivity. In ovariectomized rats, estrogen supplementation increased EDH-mediated vasodilation in uterine arteries,[47] while pregnancy and prolonged administration of 17β-estradiol to non-pregnant sheep enhanced uterine artery production of eNOS, cGMP-dependent protein kinase 1 (PKG-1α), and cGMP, activating BKCa channels through phosphorylation events.[51-53]

Vascular endothelial growth factor (VEGF) facilitates pregnancy-induced overexpression of SKCa2.3 and IKCa channels, by preventing caveolin-mediated internalization and increasing their abundance in plasma membrane. In cultured human uterine microvascular endothelial cells, the VEGF-dependent elevation in SKCa2.3 and IKCa levels induced by serum from normal pregnant women is blunted by VEGF receptor inhibition.[54] Moreover, BKCa channel function in uterine arteries may be modulated by epigenetic modifications, specifically methylation and demethylation at the promoter level, during normal and abnormal pregnancies.[55]

**Alterations of K$^+$ channels in preeclampsia**

Preeclampsia is an obstetric, multisystemic syndrome marked by hypertension and proteinuria, arising from extensive endothelial dysfunction. Approximately 5-8% of women worldwide are affected, causing more than 70,000 maternal and 500,000 infant deaths each year. Clinical presentation, severity and disease progression can be highly variable, making preeclampsia a persistent clinical challenge.

There have been significant efforts to understand the pathophysiology of preeclampsia; however, this work has not produced a clinically efficacious treatment, and delivery remains the only cure. This lack of success is often attributed to clinical subtypes, such as early (<34 weeks' gestation) vs. late-onset preeclampsia (>34 weeks' gestation), which may have distinct pathogenesis.[56] Nevertheless, maternal endothelial dysfunction and impaired placental perfusion are recognized as key features, the latter being specifically important in early-onset disease, which is also characterized by fetal growth restriction (FGR), where the fetus fails to reach its growth potential for a given gestational age. [57-59] Maternal endothelial dysfunction affects reproductive and non-reproductive vascular beds during pregnancy and persists postpartum, increasing long-term cardiovascular disease risk.[60]

A mechanistic link also exists between maternal endothelial dysfunction and placental stress in preeclampsia. Emerging evidence indicates that diminished placental perfusion, due to impaired spiral artery remodeling, or syncytiotrophoblast stress, triggers the release of proinflammatory, antiangiogenic and vasoactive factors that compromise endothelial barrier integrity, reduce vasodilatory capacity, and enhance vascular contractility.[61]

KCa channels are crucial in the context of pathological pregnancy, as highlighted in a recent review.[62] Their expression and activity decrease in preeclampsia, a change driven by reactive oxygen species, endoplasmic reticulum stress, and hypoxia-induced epigenetic regulation.[62] The loss of KCa channel function impairs transfer of endothelial cell hyperpolarization across gap junctions in human placental arteries and in the umbilical artery and vein.[63] This dysfunction extends to other vascular beds and compromises vascular health of both the mother and offspring. Maternal undernutrition, a recognized preeclampsia risk factor, also impairs EDH-mediated dilation in fetal coronary arteries when exposed to bradykinin. In these circumstances, bradykinin-induced relaxation becomes entirely NO-dependent, compensating for the loss of EDH.[64]

In preeclampsia, circulating factors that modulate EDH are significantly impacted. Maternal serum concentrations of soluble fms-like tyrosine kinase 1 (sFlt-1) increase, while VEGF and PlGF decrease. sFlt-1 antagonizes both angiogenic factors, promoting vasoconstriction and endothelial injury that drive FGR and preeclampsia.[65] Elevated sFlt1 may also induce vasoconstriction by inhibiting the VEGF-mediated reduction of SKCa and IKCa expression. Endoplasmic reticulum stress and protein kinase C further impair KCa channel function, leading to maladaptive alterations in uteroplacental circulation. [66]

**Therapeutic strategies to enhance canonical vasorelaxant signaling pathways**

Pharmacological approaches enhancing vasodilatory pathways have been actively sought to address tissue ischemia. NO-dependent vasodilation can be induced via substrates and co-factors of eNOS and, more commonly, using NO donor pharmaceuticals, sGC activators/stimulators such as riociguat, cinaciguat, and vericiguat, and phosphodiesterase inhibitors like sildenafil.[67] These drugs are effective in conduit vessels, where endogenous NO signaling is most pertinent. The indications for NO mimetics encompass angina pectoris, erectile dysfunction, pulmonary arterial hypertension, and neurological diseases.[68] However, they are unsuitable for hypertension or microangiopathic disease, conditions for which drugs that specifically target the KCa channels would be preferable.[69] A Cochrane Database Systematic Review of randomized controlled trials determined inadequate evidence to endorse using NO donors, L-arginine, or NOS inhibitors in acute stroke.[70] Similarly, while diminished eNOS is a common characteristic of preeclampsia, adverse effects, including headaches and the onset of tolerance hinder the application of NO donors in this condition.[71]

Minimal concentrations of NO, ranging from picomolar to nanomolar, facilitate vasodilation and cellular protection, whereas elevated concentrations, spanning micromolar to millimolar, engage with reactive oxygen species to generate reactive nitrogen species, resulting in cytotoxicity through the nitration and/or nitrosation of proteins, lipids, and DNA.[72] NO inhibitors may be employed experimentally to assess the significance of NO signaling in biological processes or therapeutically when excessive NO production leads to cellular stress and damage.

**Efforts to create pharmaceuticals targeting K$^+$ channels.**

There is a great interest in developing new therapeutic products targeting endothelial ion signalling to rescue blood flow in microvascular disorders. Nonetheless, no clinical drug is currently available to improve hyperpolarization-mediated vasodilation. A first class of compounds have been developed to impact hyperpolarization via the activation of KCa channels, exploiting their diverse organotypic expression to achieve organ-specific effects. For instance, BKCa channels are mainly found in smooth muscle cells, and the inner mitochondrial membrane of cardiomyocytes; activation of BKCa channels in these locations results in vasodilation and protection against cardiac ischemia.[73] IKCa channels are expressed in smooth muscle cells, endothelial cells, and cardiac fibroblasts, and therefore, their pharmacological modulation can impact vessel dilation and fibrogenic potential.[74] SKCa channels are extensively expressed in endothelial cells and the nervous system, where activation primarily leads to membrane hyperpolarization, inhibiting cell firing and reducing repetitive action potentials' frequency.[10] Positive KCa2.X/KCa3.1 channel gating modulators, such as NS309, stabilize the open conformation induced by Ca$^{2+}$-CaM binding rather than activating the KCa channels.[75] This results in a leftward shift of the concentration-response curve for KCa2.X/KCa3.1 channel opening, thereby generating a more substantial electrical response at a given concentration of cytosolic Ca$^{2+}$. First-generation compounds like NS309 and DCEBIO are unsuitable for in vivo studies due to potential off-target effects and short plasma half-lives. Consequently, their application was restricted to in vitro studies demonstrating that enhanced agonist-evoked hyperpolarization improves endothelial function.[76,77]

Wulff and colleagues have synthesized a novel series of KCa2.X/KCa3.1 channel-positive modulators using the neuroprotective agent riluzole as a chemical scaffold to generate more selective compounds and better suited for in vivo use.[78] SKA-31 is the prototype of this second-generation KCa channel modulator that acts on IK1/SK channels. The compound reportedly induces vasorelaxation of isolated resistance arteries pre-constricted with phenylephrine,[79] reduces blood pressure in normotensive and angiotensin II-induced hypertensive mice, and improves endothelium-dependent vasodilation and cardiac contractility in aging mice.[80] Acute treatment with a low concentration of SKA-31 (i.e., 0.3 µM) significantly enhanced the endothelium-dependent vasodilatory responses to acetylcholine and bradykinin in myogenically active cremaster skeletal muscle resistance arteries from a rat model of spontaneous type II diabetes, and also in isolated intra-thoracic resistance arteries from human diabetic subjects. These findings indicate that human arteries remain sensitive to KCa channel facilitation by a positive modulator even after many years of type II diabetes diagnosis and treatment.[81]

Selective blockade of IKCa prevents phenotypic changes in smooth muscle and coronary artery neointimal formation in two different models of post-angioplasty restenosis and during atherosclerosis in hypercholesterolemic mice.[82] IKCa (KCa3.1) channel inhibition may help decrease atherosclerotic plaque formation and induce beneficial monocyte polarization without altering the systemic immune response. IKCa channels are also involved in endothelial proliferation, cancerous cells, and angiogenesis.[83] Therefore, therapeutic activation of IKCa might be a double-edged sword causing unwanted detrimental effects, including acceleration of atherosclerosis, neointima formation, and tumoral angiogenesis. On the other hand, SKA-31-mediated facilitation of platelet KCa3.1 channels reportedly decreases platelet aggregation and adhesion, which may be helpful in the condition of damaged endothelium or unstable plaques.[84] Unexpectedly, SKA-31 or other KCa channel activator compounds inhibited the transition of coronary smooth muscle cells from a contractile to a dedifferentiated phenotype and decreased their proliferation in response to mitogenic stimulation by platelet-derived growth factor (PDGF).[85] It remains unknown why KCa3.1 channel activators and inhibitors can similarly suppress smooth muscle cell proliferation. Finally, SK3 channels are widely expressed, especially in the central nervous system, and the result of prolonged activation of this population of $K^+$ channels is unknown. Blockade of SKCa channels could help treat cognitive enhancement, depression, cardiac arrhythmias, and myotonic muscular dystrophy.[86]

Additional targets for pharmacologically induced vasorelaxation include $Ca^{2+}$-release activated $Ca^{2+}$ (CRAC) channels and transient receptor potential (TRP) channels. Nevertheless, the exact functions of these molecular components remain inadequately understood and underutilized in therapeutic applications. For example, TRPV4 channels are considered a promising target in cardiovascular diseases. However, GSK1016790A, a specific and potent agonist that elevates endothelial intracellular $Ca^{2+}$ concentration and induces endothelium-dependent relaxation, also leads to endothelial dysfunction, circulatory collapse, and mortality.[87] Iptakalim is a novel ATP-sensitive potassium channel (K-ATP) opener that exhibits high selectivity for the sulphonylurea receptor 2B (SUR2B)/Kir6.1 subtype. This compound lacks the adverse side effects associated with other nonspecific $K^+$ channel openers. Additionally, it induces arteriolar vasodilatation without impacting capillaries or large arteries. It is reported to ameliorate microvascular disturbances by inhibiting pericyte contraction, thereby addressing no-reflow phenomena following ischemic stroke.[88,89]

**Multitarget strategy**

The previous section highlighted the approach of developing pharmaceuticals to activate or inhibit a specific pathway. The "one drug, one target" model seeks to deliver potent and targeted therapeutic benefits while minimizing off-target side effects. However, this strategy may not be universally effective, as variations in key disease-related biological pathways across the general population could alter the pathogenic role of a particular target. In these instances, the influence of an alternative target may prevail, and such patients might derive more significant benefit from a combination therapy that concurrently targets both the primary and alternative pathways.

Pharmaceuticals that concurrently affect multiple targets across various vascular regions are more likely to manage conditions resulting from inadequate perfusion. For example, significant vasodilatory mechanisms exhibit varying importance in cerebrovascular disease depending on vessel size, with vascular hyperpolarization dominating in smaller vessels and NO in larger ones. Large artery and small artery disease can also coexist in the same patient, thus underscoring the necessity for combination therapies that influence both districts.

Among other benefits, drug associations result in fewer instances of drug resistance or evasion. As a result, medication combinations are now the norm for treating long-term illnesses like cancer, inflammatory diseases, type II diabetes, bacterial and viral infections, and asthma. Based on the connections between their targets, multi-target therapies can be divided into a number of groups. Therapeutic effects mediated by the drug combination's impact on specific signaling pathways within the same or various cell types or tissues fall into the first category. In order to facilitate action at a second target level, the second category involves using a single medication to modulate one target. Coordinated pharmacological activities at multiple sites on a single target or macromolecular complex fall under the third type. In many cases, the components of the combination are co-formulated into a single pill or injection, however these combinations are often used as co-therapy regimens.[90]

Typical examples are represented by multidrug treatment of resistant arterial hypertension,[91] a polypill that includes aspirin, angiotensin-converting–enzyme inhibitor, and statin for the secondary prevention of cardiovascular death and complications after myocardial infarction,[92] and NO-releasing statins combining 3-hydroxy-3-methylglutaryl-coenzyme A (HMG-CoA) reductase inhibition and slow NO release, that possess more robust anti-inflammatory and antiproliferative activities than the native statins.[93] Entresto, an oral combination that contains two blood pressure-lowering medications: sacubitril and valsartan. Sacubitril blocks neprilysin, an enzyme that degrades natriuretic (and other vasoactive) peptides, thereby relaxing blood vessels and promoting sodium and water excretion. Valsartan blocks the effects of the vasoconstrictor angiotensin II. Entresto is used to treat adults with heart failure to help reduce the risk of death and hospitalization, and to treat children aged 1 year and older who have symptomatic heart failure.[94]

The Lacunar Intervention Trial-2 (LACI-2) randomized clinical trial was designed to deliver escalating doses of isosorbide mononitrate, a NO donor, and Cilostazol, a phosphodiesterase enzyme (PDE3) inhibitor, which augments the prostacyclin-cAMP pathway in patients with small vessel stroke. The combination of these two drugs started months after the stroke and was given for over a year. Results indicate that the treatment reduced the risk of major vascular events and functional and cognitive outcomes compared to single-agent administration, with a good safety profile. These data support the concept that prolonged targeting multiple vasorelaxant pathways via NO and prostacyclin may improve functional stroke recovery.[95] Reinforcing the therapeutic value of augmenting the nitric oxide–prostacyclin signaling axis, Ohtuvayre® (ensifentrine)—a novel, nebulized PDE3/4 inhibitor that enhances both cGMP and cAMP signaling—was approved by the FDA in 2024 for improving pulmonary function in adults with moderate to severe chronic obstructive pulmonary disease.

While drug combinations enhance the efficacy of disease treatment, they also present several disadvantages, including more side effects, excessive potency, limited dosage flexibility, poor patient adherence—particularly with separate formulations—and elevated costs. Therefore, it is preferable to utilize a single drug with multitarget capabilities.

**A master key/factor to counteract hypoperfusion in stroke and pregnancies complicated by placental ischemia**

Enzymes are natural catalysts that accelerate physiological reactions, such as the cleavage of vasoactive peptides from a substrate. The peptide's interaction with its specific receptors activates various intracellular signaling pathways. This process facilitates intricate regulation of redundant pathways that have evolved to sustain physiological homeostasis, including adequate tissue perfusion. Using an analogy, an enzyme can be likened to the bow of a master key, while the resultant peptide serves as a blade capable of unlocking multiple locks. Herein, we present an example of a natural enzyme-peptide system that demonstrated therapeutic applicability in addressing tissue ischemia by activating multiple vasorelaxant pathways.

The serine protease tissue kallikrein (KLK1) is extensively expressed in various tissues, including the vasculature, which provides numerous protective effects against ischemia and promotes healing responses.[96] KLK1 cleaves kinin peptides from the substrate low-molecular-weight kininogen.[97] It is distinct from plasma kallikrein, another kinin-generating enzyme secreted into the bloodstream from the liver as an inactive precursor (plasma prekallikrein). Both prekallikrein and factor XII (FXII) exhibit intrinsic proteolytic activity. They can activate each other reciprocally to generate FXIIa and plasma kallikrein, a process facilitated by a negatively charged surface during blood coagulation.[98]

The expression and activity of KLK1 are reduced in patients with cardiovascular disease, and this deficit is associated with an increased risk of acute ischemic accidents like stroke.[99-101] Moreover, circulating maternal levels of KLK1 are significantly reduced in severe preeclampsia compared with mild preeclampsia and normal pregnancy, negatively correlating with blood pressure and proteinuria.[102] Reconstituting endogenous levels by exogenous administration of semipurified KLK1 prevented restenosis after stenting of severe atherosclerotic stenosis of the middle cerebral artery.[103] In addition, urine-extracted KLK1 therapy has been successfully trialled in China on several hundred thousand stroke patients.[104,105] Nonetheless, the derivative nature of this medical product poses significant challenges for regulatory approval and patient acceptance outside of China.

Rinvecalinase alpha (DM199) is a recombinant version of the natural protein KLK1. DM199-based therapy resolves the concerns of a semipurified protein. It has been titrated to maximize efficacy and safety. By restoring endogenous KLK1 levels and kinin production, DM199 activates multiple vasorelaxant pathways, all mediated by the kinin B2R. **(1)** Under homeostatic conditions, kinin generation by KLK1 and B2R expression is set up to maintain physiologic NO release through IP3-dependent $Ca^{2+}$ release from the endoplasmic reticulum. Following injury, such as brain ischemia, B2R expression is upregulated. Bradykinin stimulation generates prolonged high-output eNOS-derived NO.[106] **(2)** Bradykinin activates the synthesis of $PGI_2$.[107] An early report suggested that KLK1 can directly stimulate the release of $PGI_2$ from vascular cells independently of kinin formation but through other substance(s) produced by this serine proteinase.[108] **(3)** Bradykinin is also a potent activator of EDH signalling. The binding to the cognate B2 receptor activates phospholipase C (PLC), an enzyme that hydrolyzes phosphoinositides. This hydrolysis releases inositol trisphosphate (IP3) and diacylglycerol (DAG). In turn, IP3 triggers the release of $Ca^{2+}$ from intracellular stores, primarily the endoplasmic reticulum. The increase in intracellular $Ca^{2+}$ concentration is a key step in activating KCa channels. This mechanism involves the binding of CaM to the C-terminal domain of KCa2.X/KCa3.1 channels to induce their pore opening and subsequent $K^+$ efflux. Additionally, as shown in the coronary microcirculation, bradykinin can induce the release of a cytochrome P450-derived arachidonic acid metabolite, which exhibits EDHF-like features.[109] In the canine coronary artery, the hyperpolarization responses to bradykinin reportedly occurred at concentrations comparable to those initiating the NO-dependent component.[110] Pioneering research by Garland's team demonstrated that bradykinin is a potent inducer of conducted (retrograde) hyperpolarization-dependent vasodilation, independent of NO or $PGI_2$. When locally administered to human and porcine coronary arterioles, bradykinin evoked vasodilation that spread along the arteries with minimal decline for at least 1,000 μm from the application site. Endothelial SKCa and IKCa channels were responsible for the spread of local and conducted vasorelaxation.[111] Interestingly, despite poor contractile function, diseased small coronary arteries showed preservation of conducted dilation in response to focally applied bradykinin. These data suggest that enhancing kinin generation by exogenous KLK1 administration can be therapeutically helpful in improving microvascular blood flow. **Figure 1** illustrates using DM199-KLK1 as a master key to unlock the perfusion deficit.

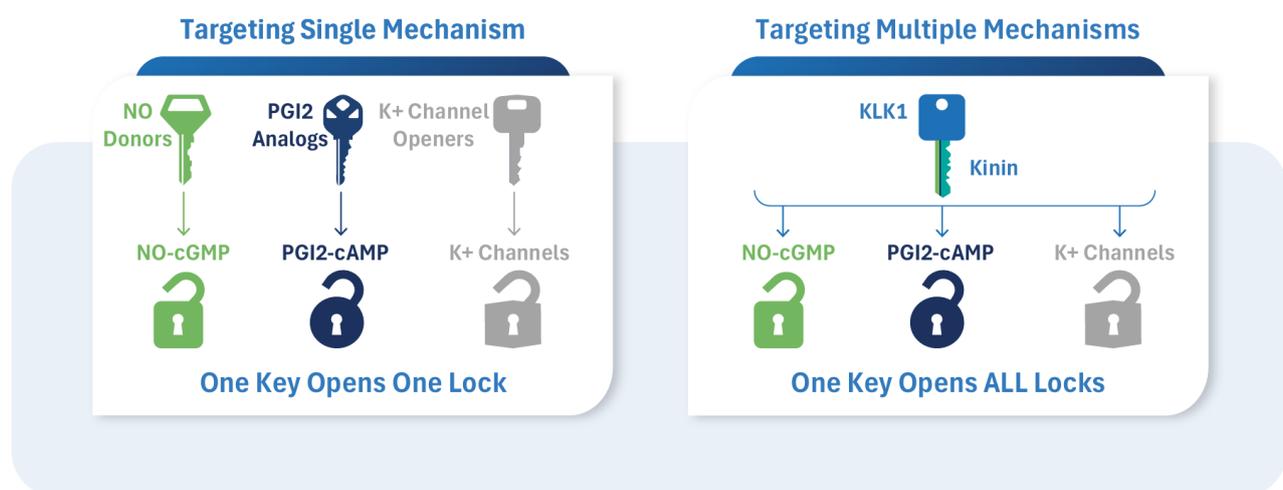

*Figure 1. A master key to activate all vasodilatory mechanisms.* *The fundamental concept of multitarget therapy posits that each pharmaceutical agent (the key) will influence a specific target (the locks) to reestablish optimal conditions (unlock the lock). DM199-based therapy is distinctive as*

*a multi-target strategy, incorporating a bow (the enzyme) and a master blade (generated kinin) that can unlock multiple locks. These attributes enhance the likelihood of clinical efficacy.*

Preliminary clinical data suggest that DM199 represents a promising novel strategy for acute ischemic stroke, noninvasively promoting early reperfusion by selectively dilating collateral arteries in the ischemic penumbra, with the potential for prolonged treatment beyond the acute phase. Significantly, with a 24-hour treatment window, DM199-based therapy may benefit numerous stroke patients who are ineligible for traditional revascularization and appears to diminish the risk of secondary strokes, as indicated by initial findings from the ReMedy1 clinical trial. DiaMedica's DM199 clinical development program expansion anticipates a Phase 2 investigator-sponsored trial focused on preeclampsia and FGR. The trial assesses whether this novel drug can effectively lower blood pressure, enhance endothelial function, and improve perfusion to maternal organs and the placenta.

**Prospective trajectories and conclusion**

In recent years, revascularization techniques through wires, catheters, and drug-eluting technology have improved the outcome of patients with large artery disease.[112] However, these procedures are not practicable in microvascular disorders, such as lacunar strokes, dementia, and preeclampsia. Mechanistic understanding of the redundant vasoactive systems has revealed that hyperpolarization currents originating from endothelial cells and propagating to surrounding mural cells play a key role in governing microvascular perfusion. Interestingly, hyperpolarization can spread bidirectionally to coopt upstream larger vessels. This $K^+$ channel-dependent conducted vasodilation appears partially conserved in microvascular disease and potentially enhanced pharmacologically as indicated by experimental studies, suggesting that hyperpolarization should be considered a pivotal therapeutic target to aid the more established treatments with NO donors. However, no clinical drug is currently available to improve hyperpolarization-mediated vasodilation. Repurposing multitask natural proteins and enzymes like KLK1 could be a viable option to simultaneously stimulate all the necessary homeostatic signalling responses, surpassing the efficacy and safety of associating different drugs. Refining multi-target pharmaceuticals may lead to personalized therapies to combat life-threatening vascular disease according to individual phenotypic variations.